\documentclass[twocolumn,description,aps]{revtex4-1}
\usepackage{graphicx,epstopdf,amsmath,amssymb,multirow,CJK,color,tabularx}
\usepackage[german,english]{babel}
\usepackage[colorlinks, linkcolor=blue,anchorcolor=blue,citecolor=blue,urlcolor=blue]{hyperref}

\begin{document}
\title{Porous-B$_{18}$: An Ideal Topological Semimetal with Symmetry-Enforced Orthogonal Nodal-Line and Nodal-Surface States}

\author{Xiao-jing Gao$^{1}$}
\author{Yanfeng Ge$^{1}$}
\author{Yan Gao$^{1}$}\email{yangao9419@ysu.edu.cn}

\affiliation{$^{1}$State Key Laboratory of Metastable Materials Science and Technology $\&$ Hebei Key Laboratory of Microstructural Material Physics, School of Science, Yanshan University, Qinhuangdao 066004, China}

\date{\today}

\begin{abstract}
Topological semimetals (TSMs) featuring symmetry-protected band degeneracies have attracted considerable attention due to their exotic quantum properties and potential applications. While nodal line (NL) and nodal surface (NS) semimetals have been extensively studied, the realization of a material where both NL and NS coexist and are intertwined, particularly with an ideal electronic band structure, remains a significant challenge. Here, we predict via first-principles calculations and symmetry analysis a metastable boron allotrope, Porous-B$_{18}$ (space group $P6_3/m$, No.~176), as a pristine TSM hosting a NS and two straight NLs near the Fermi level. The structure, a honeycomb-like porous 3D framework, exhibits excellent dynamical, thermal (stable up to 1000~K), and mechanical stability. Its electronic band structure is remarkably clean: only the highest valence band (HVB) and the lowest conduction band (LCB) cross linearly within a large energy window of 1.84~eV, free from trivial-band interference. The nodal surface lies on the $k_z = \pm \pi$ planes, protected by combined time-reversal symmetry ($T$) and twofold screw-rotational symmetry ($S_{2z}$), yielding a full-plane Kramers-like degeneracy. The two nodal lines along $K$--$H$ and $K'$--$H'$ are protected by inversion and time-reversal symmetries, carry a quantized Berry phase of $\pm \pi$, and connect orthogonally to the nodal surface, forming an intertwined nodal network. Drumhead surface states on the $(1\bar{1}0)$ surface further confirm the nontrivial topology. Porous-B$_{18}$ thus provides an ideal platform for investigating the interplay between nodal-line and nodal-surface fermions and exploring novel quantum transport phenomena.
\end{abstract}

\date{\today} \maketitle

\section{INTRODUCTION}\label{sec_introduction}

The discovery of topological phases of matter has fundamentally transformed our understanding of modern condensed matter physics, evolving from the initial focus on topological insulators~\cite{TI-Hasan,TI-TS} to encompass a rich variety of topological semimetals (TSMs)~\cite{WDSMs,TSMs,TQM-classSymm,TSMsFab}. According to the dimensionality of band degeneracy manifolds near the Fermi level ($E_{\text{F}}$), TSMs can be systematically classified into zero-dimensional (0D) nodal-point semimetals (e.g., Dirac, Weyl, and multifold degenerate points)~\cite{WanWSM,WengTaAs,YoungDSM,WangNa3Bi,TPs,UnMPBradlyn}, one-dimensional (1D) nodal-line semimetals (NLSMs)~\cite{FangNLSM,GaoNLSM,MTCsNLSM}, and two-dimensional (2D) nodal-surface semimetals (NSSMs)~\cite{YangNS,LiangNSs,ZhongNSs,ChenNS}. NLSMs typically manifest as closed loops/curves or straight lines in momentum space and are often associated with unique surface signatures~\cite{FangNLSM}, such as the drumhead-like surface states~\cite{MTCsNLSM}. NSSMs, representing the highest-dimensional band degeneracy, provide exceptional platforms for studying intriguing quantum behaviors, including enhanced quantum oscillations and exotic plasmon excitations~\cite{YangNS,ZhongNSs,LiangNSs}.

Significant advances have been made in identifying candidate materials for NL and NS semimetals. NLSMs have been experimentally or theoretically suggested in systems such as Cu$_3$PdN~\cite{Cu3PdN}, PbTaSe$_2$~\cite{PbTaSe2}, and Ca$_3$P$_2$~\cite{Ca3P2}, and various carbon allotropes~\cite{ChenYP-TCrabon,WangJT-TCarbon}. Likewise, NSSMs have been theoretically predicted in several material systems~\cite{YangNS,LiangNSs,ZhongNSs,ChenNS}. Nevertheless, fundamental challenges persist~\cite{YangNS}. In many proposed nodal-surface materials, strong spin--orbit coupling (SOC) gaps the degeneracies, undermining their topological character. Moreover, nodal features often appear far from $E_{\text{F}}$ or are obscured by trivial bands, complicating experimental detection. Crucially, topological NLs and NSs typically exist separately in different materials, hindering systematic investigation of their mutual interactions and the emergent physics arising from these distinct higher-dimensional topological fermions.

\begin{figure*}[!th]
	\centering
	\includegraphics[width=0.98\textwidth]{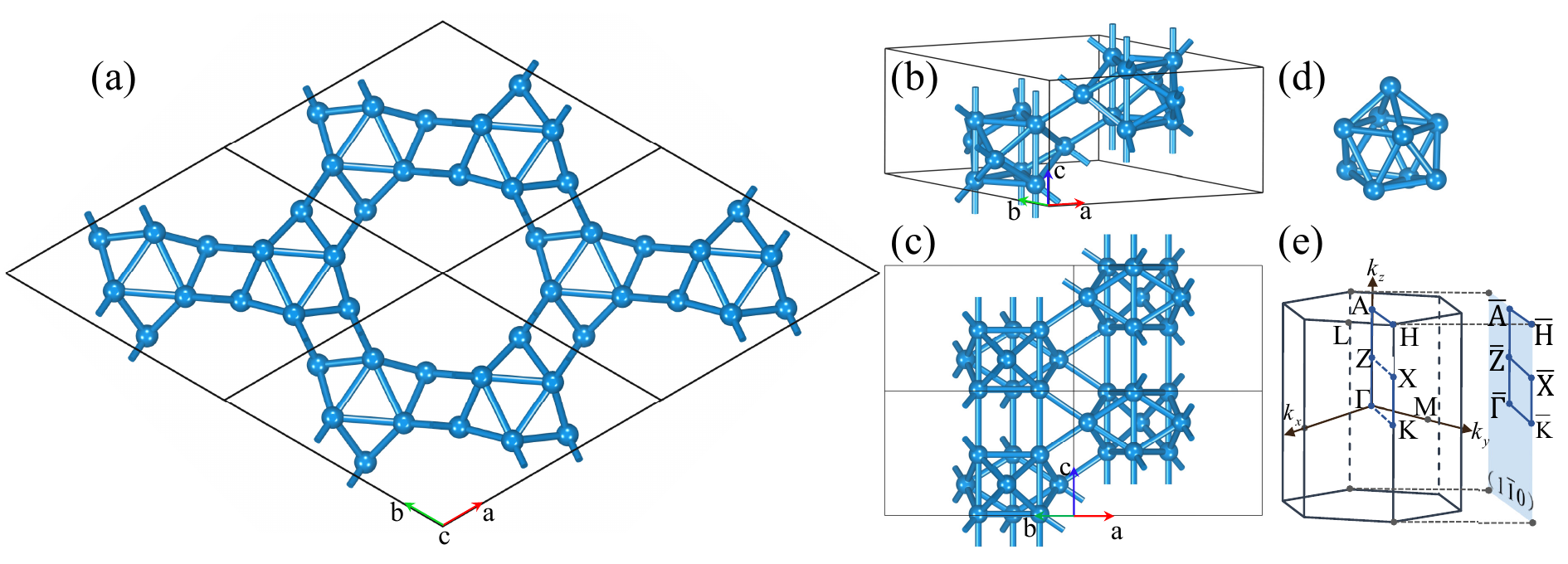}
	\caption{(Color online) Crystal structure and Brillouin zone (BZ) of Porous-B$_{18}$. (a) Top view of a $2\times2\times2$ supercell. (b) Perspective view of the primitive cell. (c) Side view of a $1\times1\times2$ supercell. (d) The B$_9$ cage as the basic structural unit. (e) The first BZ and the projected $(1\bar{1}0)$ surface BZ.}
	\label{fig_structure}
\end{figure*}

Recent theoretical works have highlighted materials such as Ti$_3$Al~\cite{Ti3Al}, NaAlSi~\cite{NaAlSi}, and $X$TiO$_2$ ($X$=Li, Na, K, Rb)~\cite{XTiO2}, in which both nodal lines and surfaces appear. However, these systems often suffer from non-negligible SOC that breaks band degeneracies, or their nodal features are located far from the $E_{\text{F}}$ and coexist with numerous trivial bands. An ideal hybrid-dimensional TSM hosting intertwined nodal lines and surfaces should therefore satisfy several criteria: (i) negligible SOC, achievable with light elements; (ii) clean, isolated crossings formed by only two bands near $E_{\text{F}}$; and (iii) robust, interconnected nodal manifolds spanning a large energy window. Such a material has remained elusive.

Boron allotropes offer a compelling materials platform in this context. The inherently weak SOC in boron preserves symmetry-enforced degeneracies, while its electron-deficient bonding supports a rich variety of metastable structures, from 0D clusters to 3D frameworks~\cite{WangReBoron,Tai2DB,YakobsonBoron,Goddard3Dboron}. Several 3D boron phases~\cite{3Dborophene,3DalphaB,H-boron,Pnma-B60,AB-16-Pnnm,Ort-B32} have been predicted to exhibit nodal points, lines, or surfaces, suggesting the possibility of more complex, hybrid topological states. This raises a natural question: can a 3D boron allotrope host coexisting and interconnected nodal lines and surfaces in an ideal electronic environment?

\begin{table*}[!th]
	\renewcommand{\thetable}{I}
	\caption{\label{tab:I} Comparison of structural parameters (space groups, lattice parameters (\AA), angles ($^\circ$), and bond lengths (\AA)), density (g/cm$^3$), bulk modulus (GPa), total energy ($E_{\text{tot}}$, eV/B), and topological properties for Porous-B$_{18}$ and other 3D boron allotropes. NL-NSSM denotes a topological semimetal characterized by the coexistence of nodal-line and nodal-surface states.}
	\centering
	\setlength{\tabcolsep}{6pt}
	\renewcommand{\arraystretch}{1.12}
	\begin{tabular*}{\textwidth}{@{\extracolsep{\fill}} l c c c c c c c c c c c c }
		\hline\hline
		\multirow{2}{*}{Structure} & \multirow{2}{*}{Space groups} & \multicolumn{3}{c}{Lattice parameters} & \multicolumn{3}{c}{Angles} & Band & \multirow{2}{*}{Density} & Bulk & \multirow{2}{*}{$E_{\text{tot}}$} & \multirow{2}{*}{Properties} \\
		\cline{3-8}
		& & a & b & c & $\alpha$ & $\beta$ & $\gamma$ & lengths & & moduluses & & \\
		\hline
		Porous-B$_{18}$ & $P6_{3}/m$ & 6.72 & 6.72 & 3.85 & 90 & 90 & 120 & 1.73--2.02 & 2.14 & 168.91 & -6.34 & NL-NSSM \\
		H-boron & $P6_{3}/mmc$ & 6.06 & 6.06 & 9.91 & 90 & 90 & 120 & 1.62--1.71 & 0.91 & 70.42 & -5.84 & NLSM \\
		Ort-B$_{32}$ & $Cmcm$ & 6.61 & 8.09 & 4.98 & 90 & 90 & 90 & 1.63--2.09 & 2.16 & 194.00 & -6.29 & NLSM \\
		3D borophene & $C2/m$ & 5.47 & 2.81 & 1.81 & 90 & 70 & 90 & 1.73--1.89 & 2.52 & 242.34 & -6.31 & NLSM \\
		3D $\alpha'$ boron & $Cmcm$ & 7.73 & 8.23 & 5.07 & 90 & 90 & 90 & 1.66--1.85 & 1.78 & 166.16 & -6.36 & NLSM \\
		AB-16-Pnnm & $Pnnm$ & 3.20 & 8.48 & 4.50 & 90 & 90 & 90 & 1.61--1.91 & 2.35 & 218.32 & -6.42 & NLSM \\
		Pnma-B$_{60}$ & $Pnma$ & 11.82 & 4.90 & 7.52 & 90 & 90 & 90 & 1.72--1.87 & 2.47 & 235.76 & -6.60 & NLSM \\
		$\gamma$-B$_{28}$ & $Pnnm$ & 5.04 & 5.61 & 6.92 & 90 & 90 & 90 & 1.66--1.90 & 2.57 & 245.22 & -6.68 & Insulator \\
		$\alpha$-B$_{12}$ & $R\overline{3}m$ & 4.89 & 4.89 & 12.55 & 58 & 58 & 58 & 1.67--1.80 & 2.48 & 238.11 & -6.71 & Insulator \\
		\hline\hline
	\end{tabular*}
\end{table*}

In this work, we predict such a system, a porous 3D boron phase termed Porous-B$_{18}$ (space group $P6_3/m$, No.~176), identified through first-principles calculations and symmetry analysis. We confirm its dynamical, thermal (up to 1000~K), and mechanical stability. Crucially, the electronic structure shows isolated crossings between the highest valence band and the lowest conduction band, forming a nodal surface on the $k_z = \pm \pi$ planes and two straight nodal lines along K--H and K$'$--H$'$, all within a 1.84~eV energy window near $E_{\text{F}}$ devoid of trivial bands. The nodal surface is protected by time-reversal symmetry and nonsymmorphic twofold screw-rotational symmetry; the nodal lines carry a quantized Berry phase and connect orthogonally to the nodal surface, forming an integrated nodal network. Calculated drumhead surface states on the $(1\bar{1}0)$ surface provide further evidence of nontrivial topology. Porous-B$_{18}$ thus represents a realization of an ideal hybrid-dimensional TSM, opening a new avenue for exploring the physics of coupled higher-dimensional fermions and offering potential applications stemming from its unique porous structure.

\begin{figure}[!th]
	\centering
	\includegraphics[width=0.42\textwidth]{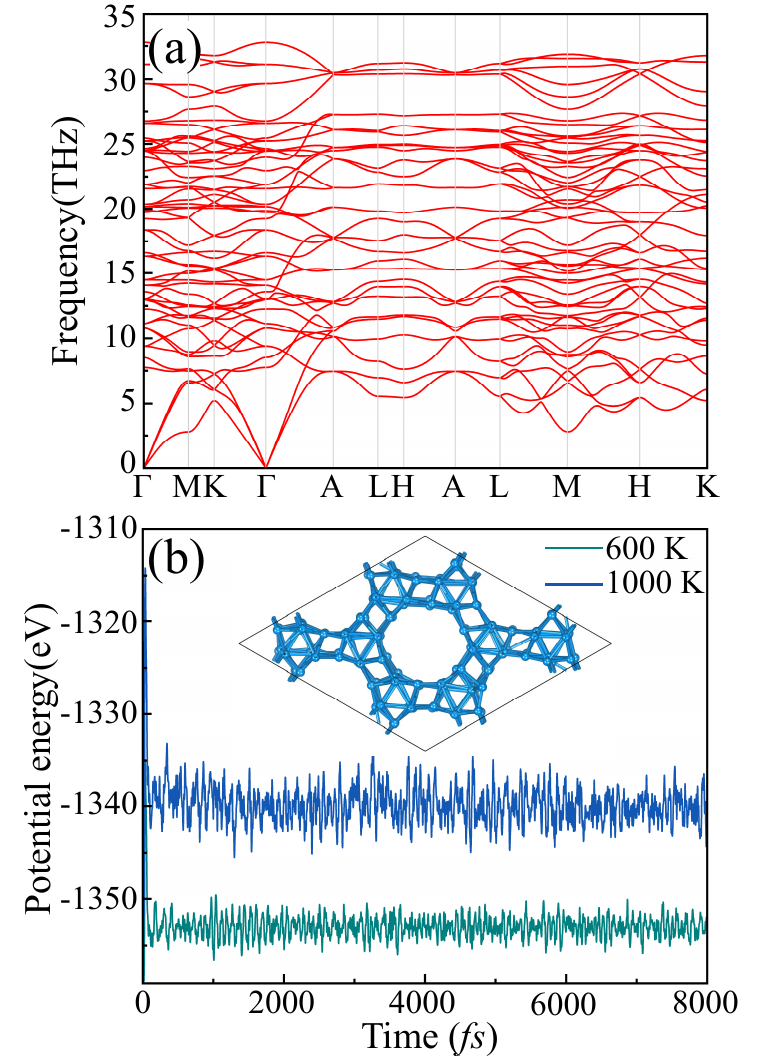}
	\caption{(Color online) Stability analysis of Porous-B$_{18}$. (a) Phonon dispersion spectrum, showing no imaginary frequencies. (b) \textit{Ab initio} molecular dynamics (AIMD) simulation results at 600~K (green) and 1000~K (blue) for 8~ps. The inset shows the snapshot of the $2\times2\times3$ supercell after 8~ps at 1000~K, which maintains its structural integrity.}
	\label{fig_stability}
\end{figure}

\section{METHOD}\label{sec_method}

Our first-principles calculations were performed within the framework of density functional theory (DFT) as implemented in the VASP package~\cite{VASP}. The generalized gradient approximation (GGA) with the Perdew--Burke--Ernzerhof (PBE) functional~\cite{PBE} was used to describe the exchange--correlation potential. The ion--electron interaction was treated using the projector augmented-wave (PAW) method~\cite{PAW}, and a plane-wave energy cutoff of 500~eV was adopted. The Brillouin zone (BZ) was sampled using a $\Gamma$-centered $k$-point mesh~\cite{k-mesh} of $10\times10\times10$ for structural optimization and electronic structure calculations. The structures were fully relaxed until the residual forces on each atom were less than 0.001~eV/\AA and the total energy convergence criterion was $1\times10^{-6}$~eV. The phonon spectrum was calculated using the finite displacement method as implemented in the Phonopy package~\cite{Phonopy}, utilizing a $2 \times 2 \times 3$ supercell. \textit{Ab initio} molecular dynamics (AIMD) simulations in the canonical ensemble with a Nos\'{e}--Hoover thermostat~\cite{AIMD} were conducted at 600~K and 1000~K for 8~ps to assess thermal stability. Elastic constants were calculated by straining the cell and fitting energy--strain~\cite{elastic-constants} relations. The topological properties and surface states were investigated using a tight-binding Hamiltonian constructed from maximally localized Wannier functions with the Wannier90 code~\cite{Wannier90}, with surface states subsequently computed via the iterative Green's function method for semi-infinite systems as implemented in WannierTools~\cite{WannierTools}.

\section{RESULTS}\label{sec_results}

\subsection{Crystal Structure and Stability}
\label{subsec:crystal_stability}

Porous-B$_{18}$ crystallizes in the hexagonal system with the centrosymmetric space group $P6_3/m$ (No.~176, point group $C_{6h}$). As shown in Figs.~\ref{fig_structure}(a)-\ref{fig_structure}(c), the primitive cell contains 18 boron atoms. The optimized lattice parameters are $a = b = 6.724$~\AA and $c = 3.855$~\AA. The boron atoms occupy two inequivalent Wyckoff positions: 12i (0.6809, 0.5030, 0.5116) and 6h (0.3535, 0.9239, 0.2500). The B--B bond lengths range from 1.73 to 2.02~\AA. The fundamental building block of the structure is a B$_9$ cage, as depicted in Fig.~\ref{fig_structure}(d). These B$_9$ cages connect to form a 3D framework. A striking feature of Porous-B$_{18}$ is its porous nature; when viewed along the $c$-axis [Fig.~\ref{fig_structure}(a)], the arrangement of B$_9$ cages forms a honeycomb-like lattice with large hexagonal pores, analogous to graphene. This porous nature results in a relatively low density of 2.14~g/cm$^3$ and a bulk modulus of 168.91~GPa, comparable to other porous boron allotropes such as Ort-B$_{32}$~\cite{Ort-B32} and 3D-$\alpha'$ boron~\cite{3DalphaB} [see Table~\ref{tab:I}]. The calculated formation energy $E_{\text{tot}} = -6.34$~eV/B suggests that Porous-B$_{18}$ is a metastable phase, but it is energetically more favorable than several previously reported NLSM boron structures like H-boron ($-5.84$~eV/B)~\cite{H-boron}, Ort-B$_{32}$ ($-6.29$~eV/B)~\cite{Ort-B32}, and 3D borophene ($-6.31$~eV/B)~\cite{3Dborophene}, and is comparable to 3D-$\alpha'$ boron ($-6.36$~eV/B)~\cite{3DalphaB}, indicating its potential for experimental synthesis.

\begin{figure*}[!th]
	\centering
	\includegraphics[width=0.86\textwidth]{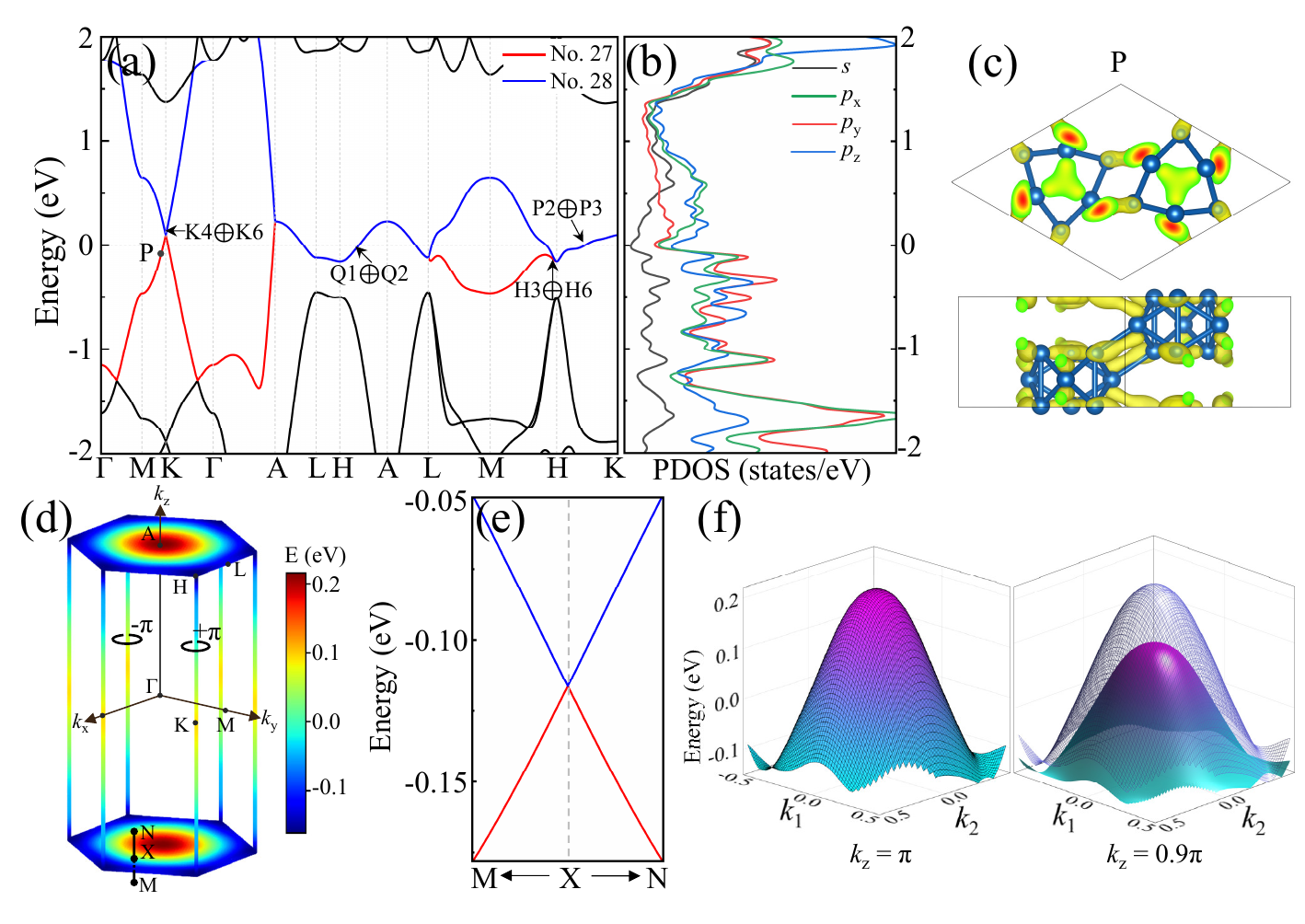}
	\caption{(Color online) Electronic properties of Porous-B$_{18}$. (a) Electronic band structure along high-symmetry paths. The irreducible representations (irreps) of the crossing bands are labeled. (b) Projected density of states (PDOS). (c) Charge density distribution of the crossing states at point P. (d) Brillouin zone showing the nodal surface on the $k_z = \pm \pi$ planes and nodal lines along the K--H and K$'$--H$'$ directions. (e) Band dispersion along the direction normal to the nodal surface at a generic point X on the nodal surface. (f) 3D band crossing visualization in the $k_z = \pi$ (left) and $k_z = 0.9\pi$ (right) planes.}
	\label{fig_band}
\end{figure*}

To assess the viability of Porous-B$_{18}$, we performed a comprehensive stability analysis. First, the phonon dispersion spectrum was calculated, as shown in Fig.~\ref{fig_stability}(a). The absence of any imaginary frequencies throughout the entire BZ confirms the dynamical stability of the structure. Second, we performed \textit{ab initio} molecular dynamics (AIMD) simulations on a $2 \times 2 \times 3$ supercell at high temperatures of 600~K and 1000~K. As shown in Fig.~\ref{fig_stability}(b), the total potential energy fluctuates around a stable average value throughout the 8~ps simulation time, and the crystal structure remains intact even at 1000~K (inset of Fig.~\ref{fig_stability}(b)), demonstrating its excellent thermal stability. Third, we calculated the elastic constants: $C_{11} = 235.63$, $C_{12} = 91.66$, $C_{13} = 143.36$, $C_{33} = 292.19$, and $C_{44} = 189.12$ (all in GPa). These values satisfy the Born stability criteria~\cite{Born-criteria} for a hexagonal system ($C_{11} > |C_{12}|$, $2C_{13}^2 < C_{33}(C_{11}+C_{12})$, and $C_{44} > 0$), confirming its mechanical stability.

\subsection{The Ideal Two-Band Crossing}

The most fascinating properties of Porous-B$_{18}$ lie in its electronic structure. Given that boron is a light element, the SOC effect is negligible, and the system can be treated as spinless. The projected density of states (PDOS) in Fig.~\ref{fig_band}(b) shows that the states near the Fermi level are predominantly contributed by boron $p$ orbitals, in agreement with the charge density at the point P [Fig.~\ref{fig_band}(c)]. The DOS diminishes at the Fermi level, which is a hallmark of a semimetal. The electronic states near the Fermi level are defined strictly by the crossing between the highest valence band (HVB, 27th band) and the lowest conduction band (LCB, 28th band) [see Fig.~\ref{fig_band}(a)]. This two-band crossing is isolated, with no other trivial bands interfering within an exceptional energy window spanning from $-0.47$~eV to $1.37$~eV. This $1.84$~eV range of isolation is critical, guaranteeing that the topological phenomena near $E_{\text{F}}$ can be analyzed exclusively through a minimal two-band model, greatly simplifying analytical theory and experimental interpretation. Within this clean window, the topological nodes are robustly positioned around $E_{\text{F}}$: the NS exists between $-0.16$~eV and $0.23$~eV, and the NLs exist between $-0.16$~eV and $0.10$~eV. The observed linear dispersion along paths such as M--K--$\Gamma$ and $\Gamma$--A [Fig.~\ref{fig_band}(a)] confirms the Dirac/Weyl-like nature of these band crossings along their normal directions. Porous-B$_{18}$ thus realizes a nearly perfect scenario for investigating distinct 1D and 2D degenerate fermions.

\begin{figure}[th]
	\centering
	\includegraphics[width=0.41\textwidth]{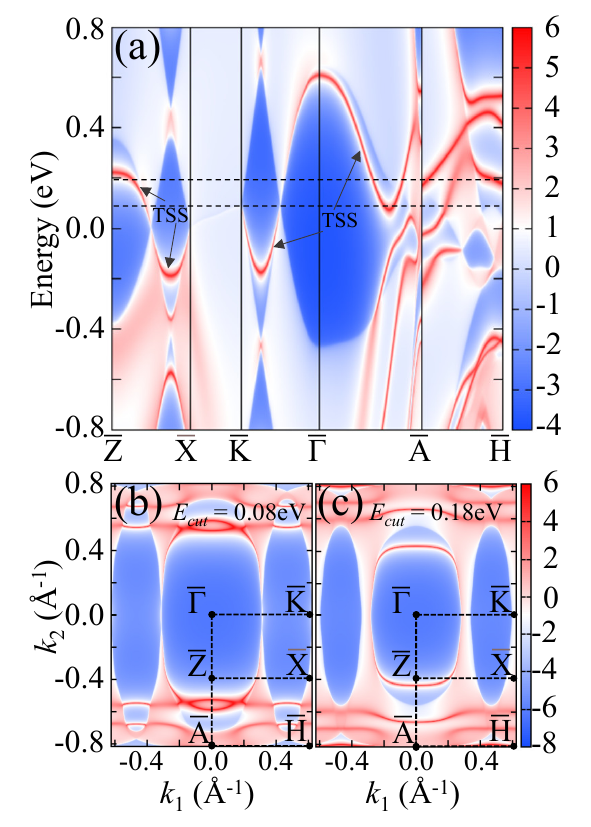}
	\caption{(Color online) Surface states on the $(1\bar{1}0)$ surface. (a) Surface band structure along the high-symmetry paths. The drumhead surface states (TSS) are highlighted. Constant energy contours at (b) $E = 0.08$~eV and (c) $E = 0.18$~eV, respectively, showing the surface states confined between the projections of the bulk nodal lines.}
	\label{fig_surfaces}
\end{figure}

\subsection{Symmetry-Enforced Nodal Surface}

A full scan of the BZ for the crossings between the HVB and LCB reveals a hybrid-dimensional topological structure, as schematically shown in Fig.~\ref{fig_band}(d). The crossings form two distinct topological elements: a 2D nodal surface (NS) and two 1D nodal lines (NLs). The NS lies on the $k_z = \pm \pi$ planes (i.e., the $A$--$L$--$H$ plane of the BZ). This 2D degeneracy is enforced by the combination of time-reversal symmetry ($\mathcal{T}$) and a nonsymmorphic two-fold screw-rotation symmetry ($S_{2z} = \{C_{2z}|00\frac{1}{2}\}$) inherent in the $P6_3/m$ space group. The stabilization mechanism relies on the anti-unitary operator $\mathcal{A} = \mathcal{T}S_{2z}$. The $k_z = \pm \pi$ plane is invariant under this operator $\mathcal{A}$. In the absence of SOC with time-reversal symmetry $\mathcal{T}$, since $[S_{2z},\mathcal{T}]=0$, the square of this combined symmetry is $\mathcal{A}^2=(S_{2z}\mathcal{T})^2=S_{2z}^2\mathcal{T}^2$. As $\mathcal{T}^2=1$ for spinless fermions and $S_{2z}^2=\{C_{2z}^2|001\}=\{E|001\}=T_{001}$ (a full lattice translation along $c$), the action of $\mathcal{A}^2$ on a Bloch state $\psi_k$ is: $\mathcal{A}^2\psi_k=T_c\psi_k=e^{-i\mathbf{k}\cdot\mathbf{c}}\psi_k=e^{-ik_z c}\psi_k$ (We use the convention $k_z=\pi/c$ for the BZ boundary). For any $k$-point on the $k_z=\pi$ plane (i.e., $k_z c=\pi$), the eigenvalue of the $\mathcal{A}^2=e^{-i\pi}=-1$ (for all $\mathbf{k}\in k_z=\pi$ plane). This is an anti-unitary symmetry that squares to $-1$. Akin to the Kramers theorem for spinful systems, this ``spinless'' or nonsymmorphic Kramers degeneracy forces every band to be at least two-fold degenerate at every single point on the $k_z=\pi$ BZ boundary.

This symmetry argument proves that the degeneracy of HVB and LCB on the $k_z=\pi$ plane is not accidental but is symmetry-enforced, forming a 2D nodal surface. Our DFT calculations perfectly confirm this. Fig.~\ref{fig_band}(f) (left panel) shows the 3D band structure in the $k_z=\pi$ plane, where the HVB and LCB are completely degenerate. As soon as we move off this plane (e.g., to $k_z=0.9\pi$, Fig.~\ref{fig_band}(f) right panel), the degeneracy is lifted, and a gap opens. Fig.~\ref{fig_band}(e) shows the dispersion along a path $M$--$X$--$N$ normal to the NS, revealing a clear linear crossing at the $X$ point (which lies on the NS), confirming the nodal character.

\subsection{Straight Nodal Lines}

In addition to the NS, the band structure in Fig.~\ref{fig_band}(a) shows a robust two-fold degeneracy along the entire K--H and K$'$--H$'$ high-symmetry lines. These 1D straight nodal lines (NLs) are also topologically nontrivial. In a system with $\mathcal{PT}$ symmetry (both inversion $\mathcal{P}$ and time-reversal $\mathcal{T}$ are present), 1D NLs can be characterized by a quantized 1D winding number (or Berry phase). To confirm their nontrivial topology, we calculated the quantized 1D winding number $\nu$ for a loop enclosing each NL, defined as $\nu = (1/\pi) \oint \mathcal{A}(\mathbf{k}) \cdot d\mathbf{k}$, where $\mathcal{A}(\mathbf{k})$ is the Berry connection ($\mathcal{A}(\mathbf{k})=i\langle u_{\mathbf{k}}|\nabla_{\mathbf{k}}|u_{\mathbf{k}}\rangle$). Our calculation yields a quantized winding number of $\nu=+1$ for the K--H line and $\nu=-1$ for the K$'$--H$'$ line [see Fig.~\ref{fig_band}(d)]. This non-zero integer invariant confirms that these are topologically stable nodal lines that cannot be gapped without breaking the protecting symmetries. 

The simultaneous existence and geometric configuration of the NLs and NSs are scientifically significant. The vertical NLs (K--H and K$'$--H$'$) intersect the horizontal NS planes ($k_z = \pm \pi$). This direct intersection in momentum space provides a rare scenario for studying the topological coupling between 1D and 2D fermions, contrasting with Ti$_3$Al~\cite{Ti3Al}, NaAlSi~\cite{NaAlSi}, and $X$TiO$_2$ ($X$=Li, Na, K, Rb)~\cite{XTiO2} structures where such features are spatially separated (e.g., one at $k_z = 0$ and the other at $k_z = \pi$).

\subsection{Topological Surface States}

A key signature of a NLSM is the presence of drumhead-like surface states. The non-zero winding numbers of the K--H and K$'$--H$'$ lines mandate the existence of such states. To verify this, we calculate the surface spectral function for the $(1\bar{1}0)$ surface [Fig.~\ref{fig_structure}(e)]. The resulting surface band structure is plotted in Fig.~\ref{fig_surfaces}(a). We observe prominent topological surface states, indicated by arrows, detached from the bulk band projections. They span the regions between the projections of the bulk nodal lines [such as $\bar{Z}$--$\bar{X}$--$\bar{K}$--\textit{$\bar{\Gamma}$}--$\bar{A}$ regions in Fig.~\ref{fig_surfaces}(a)], precisely as expected for drumhead states.

The constant energy contours in Figs.~\ref{fig_surfaces}(b) and~\ref{fig_surfaces}(c) at $E = 0.08$~eV and $E = 0.18$~eV further visualize these surface states, which form a rectangular-like region around the $\Gamma$ point, enclosed by the projections of the bulk nodal lines. The existence of these drumhead surface states provides smoking-gun evidence for the nontrivial topology of the nodal lines in Porous-B$_{18}$.

\section{DISCUSSION AND CONCLUSION}\label{sec_discussion}

The discovery of Porous-B$_{18}$ presents a significant advance in the field of topological materials. Its properties distinguish it from previously reported TSMs (e.g., Ti$_3$Al~\cite{Ti3Al}, NaAlSi~\cite{NaAlSi}, and $X$TiO$_2$ ($X$=Li, Na, K, Rb)~\cite{XTiO2}), in several crucial aspects. First and foremost, it provides a rare and ideal platform for studying the coexistence and interconnection of topological features with different dimensionalities. Unlike in Ti$_3$Al~\cite{Ti3Al}, NaAlSi~\cite{NaAlSi}, and $X$TiO$_2$~\cite{XTiO2}, where the nodal loop and nodal surface are spatially separated in momentum space, the NLs and NSs in Porous-B$_{18}$ are directly connected at the H (H$'$) points of the BZ. This interconnectedness could lead to novel physical phenomena arising from the interplay between 1D NL fermions and 2D NS fermions, a topic of great theoretical interest that has lacked a suitable material host.

Second, the system is exceptionally ``clean''. As evident in Fig.~\ref{fig_band}(a), the Fermi level and the entire energy window surrounding the topological crossings (from $-0.47$~eV to $1.37$~eV) are free of any other bands. This is in stark contrast to many known TSMs, where trivial bands often clutter the Fermi surface and obscure the topological features. The ``clean'' electronic structure of Porous-B$_{18}$, combined with the fact that the nodal features cross the Fermi level, makes it an excellent candidate for experimental verification via techniques like angle-resolved photoemission spectroscopy (ARPES) and for probing its unique transport signatures.

Third, as a light-element boron allotrope, Porous-B$_{18}$ is free from the complexities of SOC. This ensures that the predicted nodal features are robust and not gapped out, unlike in many heavy-element TSMs. This makes it a pristine system for studying the fundamental physics of symmetry-protected band crossings. Furthermore, its confirmed stability is noteworthy. While many theoretically predicted allotropes are energetically unfavorable, Porous-B$_{18}$ is more stable than several other proposed topological boron phases and exhibits remarkable thermal stability up to 1000~K, suggesting that its synthesis, while challenging, may be feasible.

Finally, the unique physical structure of Porous-B$_{18}$ imparts additional potential for applications. Its honeycomb-like porous framework provides a large surface area, which is highly desirable for catalysis. The open channels could also facilitate ion transport, making it a candidate for battery electrode materials or for applications in energy storage and gas separation. The combination of these fascinating structural properties with its novel topological electronic states makes Porous-B$_{18}$ a truly multifunctional quantum material.

In conclusion, we have predicted a new stable 3D boron allotrope, Porous-B$_{18}$, using first-principles calculations. We have systematically verified its dynamical, thermal, and mechanical stability. Our analysis of its electronic structure reveals that Porous-B$_{18}$ is a rare and ideal TSM hosting coexisting and interconnected NLs and NSs. These hybrid-dimensional topological features are formed by a clean two-band crossing near the Fermi level within a large $1.84$~eV window, protected by crystal symmetries. The straight nodal lines along the K--H (K$'$--H$'$) directions are characterized by a nontrivial winding number and give rise to prominent drumhead surface states. The NSs on the $k_z = \pm \pi$ planes are protected by a combination of time-reversal and nonsymmorphic twofold screw-rotational symmetry. Our work not only introduces a new stable boron allotrope but, more importantly, provides an ideal material platform to investigate the fundamental physics of the interplay between topological states of different dimensions. The unique porous structure of Porous-B$_{18}$ further suggests its potential for a wide range of applications, paving the way for the design and discovery of new multifunctional topological quantum materials.

\begin{acknowledgments}

This work was supported by the National Natural Science Foundation of China (Grants No.~12304202), Hebei Natural Science Foundation (Grant No.~A2023203007), Science Research Project of Hebei Education Department (Grant No.~BJK2024085), and Cultivation Project for Basic Research and Innovation of Yanshan University (No.~2022LGZD001).

\end{acknowledgments}

\end{document}